# Characterization of Pd/Y multilayers with B$_4$C barrier layers using GIXR and x-ray standing wave enhanced HAXPES

M.-Y. Wu[a], Q.-S. Huang[*,b], K. Le Guen[a], V. Ilakovac[a,c], B.-X. Li[a,d], Z.-S. Wang[b], A. Giglia[e], J.-P. Rueff[a,f], P. Jonnard[*,a]

[a] Sorbonne Université, Faculté des Sciences et Ingénierie, UMR CNRS, Laboratoire de Chimie Physique - Matière et Rayonnement, boîte courrier 1140, 4 place Jussieu F-75252 Paris cedex 05, France

[b] Key Laboratory of Advanced Micro-Structured Materials MOE, Institute of Precision Optical Engineering, School of Physics Science and Engineering, Tongji University, Shanghai 200092, P. R. China

[c] Université de Cergy-Pontoise, F-95031 Cergy-Pontoise, France

[d] Northwestern Polytechnical University, 710072 Xi'an, China

[e] CNR, Istituto Officina Materiali, 34149 Trieste, Italy

[f] Synchrotron SOLEIL, L'Orme des Merisiers, Saint-Aubin, Boîte Postale 48, 91192 Gif-sur-Yvette Cedex, France

## ABSTRACT

Pd/Y multilayers are high reflectance mirrors designed to work in the 7.5-11 nm wavelength range. Samples, prepared by magnetron sputtering, are deposited with or without B$_4$C barrier layers located at the interfaces of Pd and Y layers to reduce interdiffusion, which is expected by calculating mixing enthalpy of Pd and Y. Grazing incident x-ray reflectometry is used to characterize these multilayers. B$_4$C barrier layers are found effective on reducing the Pd-Y interdiffusion. Details of the composition of the multilayers are revealed by hard x-ray photoemission spectroscopy under x-ray standing waves effect. It consists in measuring the photoemission intensity from samples that perform an angular scan in the region corresponding to the multilayer period and the incident photon energy according to the Bragg law. The experimental result indicates that Pd does not chemically react with B nor C at the Pd-B$_4$C interfaces while Y does at the Y-B$_4$C interfaces. The formation of Y-B or Y-C chemical compound can be the reason why the interfaces are stabilized. By comparing the experimentally obtained angular variation of the characteristic photoemission with the theoretical calculation, the depth distribution of each component element can be interpreted.

**Keywords:** Pd/Y, nanometric multilayers, hard x-ray photoemission spectroscopy, x-ray standing waves.

*corresponding author: philippe.jonnard@upmc.fr; huangqs@tongji.edu.cn

## 1. INTRODUCTION

Periodic multilayer mirrors are important optical components which are applied for the x-ray and extreme ultraviolet spectra ranges. The understanding of the relation between the structures of multilayer optics and their optical properties is crucial for their design. The exploration of the structures of the multilayers by multiple analyzing methods, such as reflectance and photoemission measurements, helps improve the deposition process and eventually the optical performance.

The Pd/Y multilayer was first proposed and studied by Montcalm *et al.* [1]. Such material combination is promising as simulation gives an up to 65% reflectance of such design for radiation with 9.5 nm wavelength. The potential applications vary from EUV spectroscopy, plasma diagnosis to synchrotron radiation or free electron laser instruments. A crucial factor is found that may compromise the performance of the mirror: the interdiffusion of Pd and Y. Nitridation of the Pd/Y multilayer by introducing nitrogen during the deposition process can reduce the interdiffusion [2,3]. Yet the nitridated multilayers suffer a reflectance loss. The optical performance of the mirror is thus lower than expected from theoretical calculations. Inserting $B_4C$ barrier layers at the interfaces is reported by Windt as another effective means to obtain smoother interfaces and thus improve the optical performance of the Pd/Y system [4].

In this paper we focus on the Pd/Y multilayers inserted with $B_4C$ barrier layers trying to find out more details about the mechanism of how the barrier layers reduce the interdiffusion. We report the characterization of the samples using grazing incident x-ray reflectometry (GIXR) and x-ray standing wave enhanced hard x-ray photoelectron spectroscopy (HAXPES). The experimental performances of Pd/Y multilayer mirrors are far worse from the simulations. This could be explained by many factors such as interface roughness and Pd-Y interdiffusion. We starts to find a reason for the interdiffusion by performing mixing enthalpy calculations.

HAXPES [5,6] provides a much higher value for the probed depth compared to the conventional XPS, thanks to the long inelastic mean free path (IMFP) of the emitted electrons due to their high kinetic energy. Thus HAXPES is suitable to analyse a considerable part of structure buried under the surface. The disadvantage, being the low efficiency due to the high energy of the incident photons which leads to low ionization cross section, can be counterbalanced by using high brilliance synchrotron radiation light source. X-ray standing waves (XSW), which appear as a quasi-sinusoidal periodic electric field perpendicular to the surface of the multilayer, are generated by the interference between the incident photon beam and the one reflected by the irradiated multilayer [7]. A depth distribution of the electric field intensity in the multilayer results in a depth distribution of the ionization rate. Such distribution can be modulated by varying the grazing incident angle of the photon beam. The effect can be observed by measuring the related phenomena such as characteristic x-ray emission [8,9] and photoemission [10] as a function of the incident angle. The combination of HAXPES and XSW brings unique advantages of a non-destructive characterization method with which detailed information on the multilayer structure can be obtained such as chemical compound formed at the interfaces.

## 2. THEORETICAL AND EXPERIMENTAL METHODS

The mixing enthalpy calculated following the Miedema model [11] indicates that intermixing is possible between Pd and Y layers. Pd-Y binary phase diagram [11] also shows that multiple possible compounds may form. Unfortunately the constants for the calculations concerning $B_4C$ are not available in the literature to our knowledge, thus the intermixing of Pd-$B_4C$ and Y-$B_4C$ cannot be predicted with mixing enthalpy. However the enthalpy of formation $\Delta H_f$ of palladium as well as yttrium borides and carbides can be found in the literature [1,12]. For $B_4C$/Pd the enthalpy of formation is positive, indicating a low probability of the formation of chemical compound. However for Y, $\Delta H_f$ of $YB_2$, $YB_4$, $Y_2C$ and $Y_2C_3$ are -36, -52, -32 and -51 respectively in kJ/mol. The chemical reactions of Y-B and Y-C are then expected. Yet seen the values, the chemical selectivity of the reactions (Y-B or Y-C) cannot be predicted. The $\Delta H_f$ of Y/Pd is found to be -94 kJ/mol as another evidence of the Pd-Y interdiffusion, in agreement with our calculation of mixing enthalpy with different Pd-Y mole fractions.

A series of Pd/Y-based samples is deposited using DC magnetron sputtering technique. The original design for the mirror is Pd/Y repeating 40 times and deposited on sliced and polished Si (100) wafers with a 4 nm period thickness (2 nm Pd and 2 nm Y). A 2.5 nm $B_4C$ capping layer is deposited on top of the sample to prevent the oxidation of the component metals. Three other samples are prepared: 1 nm thick $B_4C$ barrier layers are inserted at either interface of these two metals or even both interfaces to prevent the interdiffusion of the two metals. Thus the new structures of the samples are $B_4C$/Pd/Y (1/2/2 nm), Pd/$B_4C$/Y (2/1/2 nm) and $B_4C$/Pd/$B_4C$/Y (1/2/1/2 nm). The order of the layer is from the top to the bottom of the stack, so $B_4C$/Pd/Y means $B_4C$-on-Pd, then Pd-on-Y, Y-on-$B_4C$ and so on. Considering the x-ray attenuation and the IMFP of the emitted photoelectrons, we grow only 20-period structures for the last sample, which is thicker than others, instead of standard 40-period original structures.

Each sample is characterized by GIXR using Cu Kα radiation (8048 eV). The structure is determined by fitting the GIXR result using the software IMD [13]. The parameters of this structure are then introduced into the software YXRO [14] in order to anticipate the x-ray standing wave field forming inside the stack and the HAXPES result.

HAXPES measurements are performed at the GALAXIES beamline of SOLEIL synchrotron facility [15]. The incident photon energy is set to 10 keV. With such energy we can expect an IMFP of

6.4-8.5 nm of the emitted photoelectron depending on the element and the core level. This allows a probed depth, estimated to be three times of the IMFP, to be approximately equal to 4-5 periods of the multilayer. However the photoionization cross section is very small with 10 keV photons, for example $6.5\times10^{-25}$ m$^2$ for Pd $2p_{3/2}$ core level [16]. This experimental difficulty is overcome by the high flux synchrotron radiation which guarantees good quality data.

The experimental setup is presented in Figure 1(A) where the photon beam impinges onto the sample with the grazing incident angle θ recorded by a goniometer with an angular resolution approximately equal to 0.008°. The calibration of the binding energy of the photoemission spectra is carried out with the Au $4f_{7/2}$ peak which is 84.0 eV. According to the structure of the sample $B_4C/Pd/B_4C/Y$ determined by GIXR, with the incident photon energy of 10 keV, a 74% reflectance of the incident radiation can be expected at the first order Bragg angle. This indicates an intense x-ray standing wave field, leading to a clear depth distribution of the ionization rate as presented in Figure 1(B) for the $B_4C/Pd/B_4C/Y$ sample where the enhanced ionization can be found at the anti-nodal plans of the field while the reduced ionization can be found at the nodal plans in the case that the incident angle is set at the Bragg angle (0.63°). The standing wave field fades while the incident angle moves away from the Bragg angle (0.9° for example) due to the loss of reflectance. The electron analyzer is positioned perpendicular to the incident photon beam. Photoemission is recorded while the sample is rotated around the Bragg angle. A variation of the intensity of the photoemission is observed owing to the modulation of the intensity of the x-ray standing wave field.

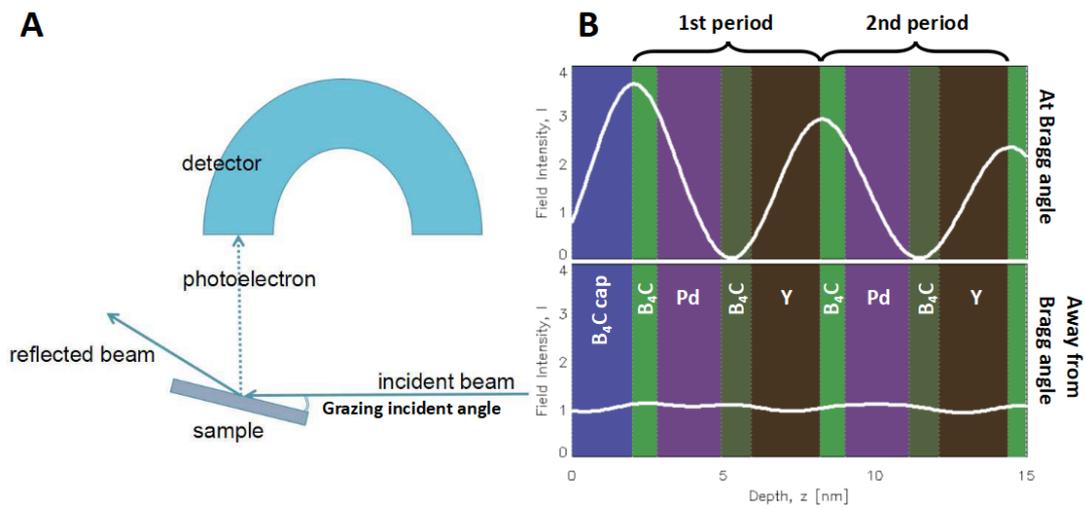

Figure 1. A: Scheme of the experimental setup. B: calculation of the depth distribution of the x-ray standing wave electric field within the $B_4C/Pd/B_4C/Y$ multilayer (6.2 nm period), with the incident beam (10 keV) introduced at first order Bragg angle (0.63°) and away from Bragg angle (0.9°).

## 3. RESULTS AND DISCUSSION

The experimental and fitted GIXR curves are presented in Figure 2. The parameters, *i.e.* thickness and interface width of the various layers, of the samples are then extracted from the fitting procedure and are listed in Table 1. To distinguish the samples easily, in the following we use sample numbers as indicated in Table 1. The interface width parameter in this table stands for both the geometrical roughness and the interdiffusion of the materials at the interfaces between the described layer and the previous one. For a Pd/Y multilayer modeled with perfect interfaces (no interface roughness nor interdiffusion), we expect a high reflectance as presented in Figure 2(A) (dotted line). However the interdiffusion, as predicted, is so severe that we barely observe the reflectance peak at the first order Bragg angle even seeing the curve in logarithm scale. The value of the interface width is almost as high as the layer thickness itself and the interferential contrast between different layers is almost completely lost. With $B_4C$ layers as barriers, clear reflectance patterns are observed for the other samples as presented in Figure 2. The interdiffusion is obviously reduced as we can read from Table 1. The fitting results show that the interdiffusion at Y-on-Pd interfaces (1.23 nm, sample 4) is stronger than the one at Pd-on-Y interfaces (0.80 nm, sample 3).

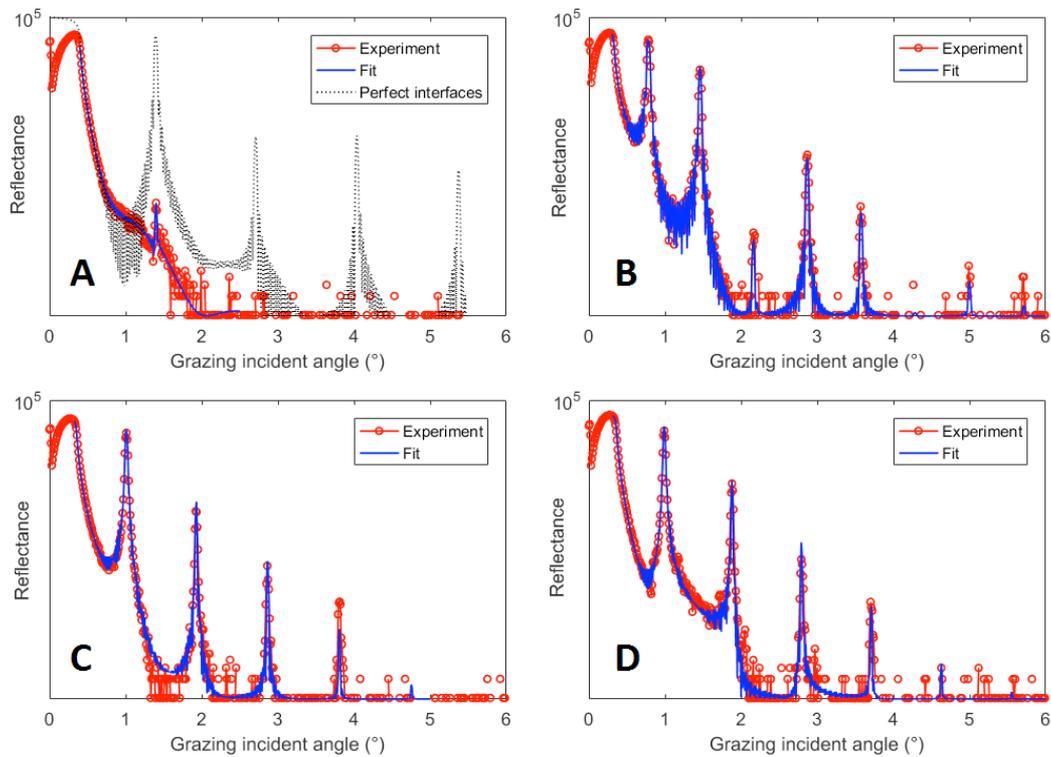

Figure 2. Measured (red dots) and fitted (blue lines) GIXR curves of the multilayers (logarithm scale). A: sample 1 (Pd/Y); B: sample 2 ($B_4C$/Pd/ $B_4C$/Y); C: sample 3 ($B_4C$/Pd/Y); D: sample 4 (Pd/ $B_4C$/Y). The dotted line in A presents the calculated reflectance of the original design without interface roughness nor interdiffusion.

| Sample | Structure | Designed thickness (nm) | Thickness (interface width) (nm) extracted from the fit of GIXR curves |
|---|---|---|---|
| 1 | $[Pd/Y]_{40}$ | 2/2 | **1.87** (1.67) / **1.43** (1.35) |

| 2 | [B$_4$C/Pd/B$_4$C/Y]$_{20}$ | 1/2/1/2 | **0.99** (0.35) / **2.10** (0.26) / **0.83** (0.27) / **2.29** (0.55) |
| 3 | [B$_4$C/Pd/Y]$_{40}$ | 1/2/2 | **0.86** (0.34) / **2.30** (0.32) / **1.50** (0.80) |
| 4 | [Pd/B$_4$C/Y]$_{40}$ | 2/1/2 | **2.19** (1.23) / **1.02** (0.30) / **1.57** (0.40) |

Table 1. Designed structures and values of the parameters extracted from the fit of the experimental GIXR curves of the samples.

The angular variations of B 1s, Pd 2p$_{3/2}$ and Y 2p$_{3/2}$ photoemission spectra of the sample B$_4$C/Pd/B$_4$C/Y are presented in Figure 3(A)-(C). The B 1s spectra photoemission are rather noisy due to the much lower ionization cross section compared to the ones of Pd 2p$_{3/2}$ and Y 2p$_{3/2}$. The so-called HAXPES-XSW curve of a core level peak is obtained in the following way: first for each of the scanned grazing angles, the intensity of the core level peak is integrated; second this integral is plotted as a function of the angle. This is repeated for all the considered core levels and the obtained result is shown in Figure 3(D). The integration is made by considering data 25 eV around the maximum of the photoemission peak. This integration range is chosen so that the background far from the maximum is not taken into account, as it may introduce some noise in the XSW curves. The integration also takes into account of the subtraction of a Shirley background [17]. The photoemission spectra of Pd 3d and Y 3d core levels are also recorded at the same time but are not presented here because their HAXPES-XSW curves has approximately the same shape of the ones of 2p and will not bring additional information. There is ignorable minor difference which is due to the incertitude or acquisition statistic because of the low intensities of 3d core level peaks. Neither are the spectra of C 1s presented, owing to the surface contamination, as the samples are preserved in the atmosphere environment.

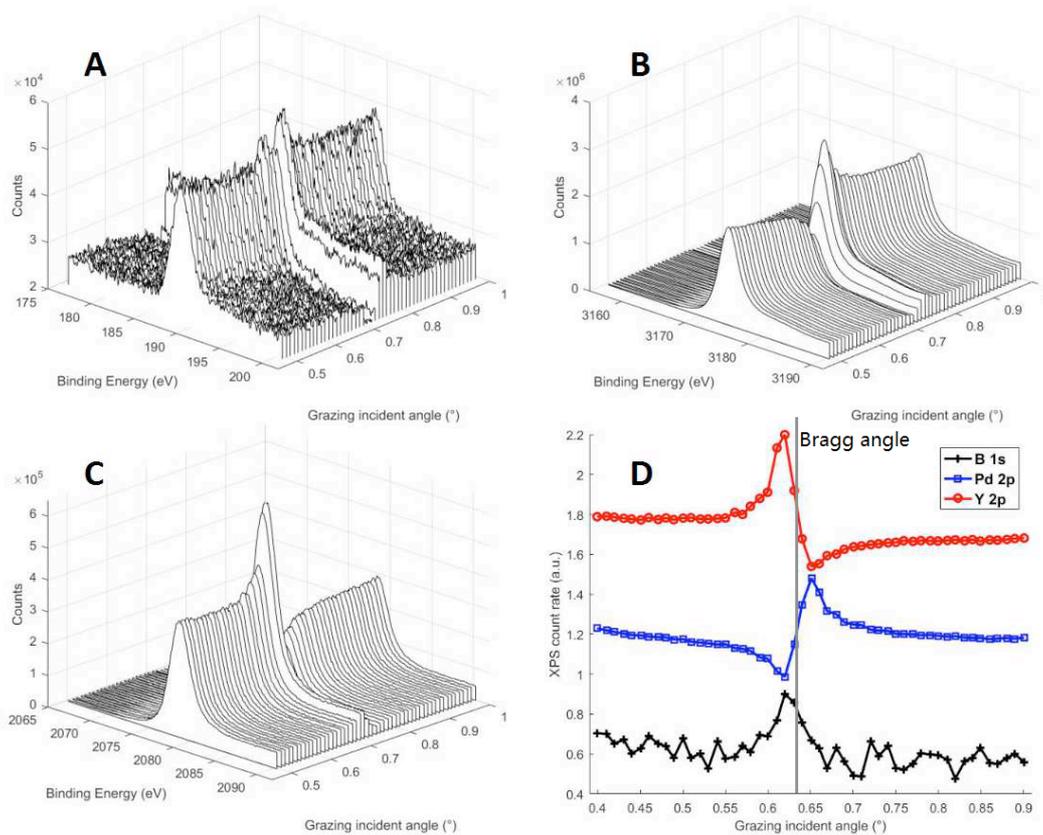

Figure 3. Photoemission spectra of the $B_4C/Pd/B_4C/Y$ multilayer as a function of the grazing incident angle in the vicinity of the Bragg angle (grey solid line) for different core levels. A: B 1s, B: Pd $2p_{3/2}$, C: Y $2p_{3/2}$, D: the corresponding HAXPES-XSW curves.

The photoemission peaks are decomposed in order to obtain the depth distribution of different chemical states for each element. We are able to retrieve some information from the Pd $2p_{3/2}$ and Y $2p_{3/2}$ spectra whose quality is reliable. Pd $2p_{3/2}$ photoemission of sample 2 is presented in Figure 4(A) where four Voigt peaks are used to fit the experimental curve, which is the sum of the spectra of all angular values in order to gain precision. For the Voigt function, the Gaussian width is 2.00 eV, which is estimated from the bandwidth of the incident photon beam; the Lorentzian width is 2.05 eV, taken from the literature [18]. The Pd metal peak is found at 3175.2 eV in binding energy. We consider other peaks with higher binding energies as satellite peaks, as were reported by de Siervo [19]. The HAXPES-XSW curve related to each Pd $2p_{3/2}$ component (metal and satellites) is plotted in Figure 4(B). The angular dependent variations of the intensities of all the satellite peaks are superposed to the one of metal peak, indicating an identical depth distribution. We have considered the possibility that the peak located at 3176.7 eV (blue solid line in Figure 4(A), main contribution beside the Pd metal) belongs to the Pd oxide. However, it is very unlikely that the oxide have the same depth distribution as the Pd metal. Indeed, the deposition of the samples is performed in pure argon. Thus the oxidation of the sample, if it penetrates the 2.5 nm $B_4C$ capping layer and enters into the multilayer, may only happen from the surface. In this case, it would be expected to have a pronounced attenuation in the depth distribution. The observation of the identical depth distribution of each contribution of the photoemission spectrum is an evidence for the assumption that the other peaks are but satellites peaks. Since only one chemical state (metal) of Pd is found in the multilayer, the diffusion of Pd and $B_4C$ layers does not form new chemical compound (Pd-B or Pd-C). The prediction by the positive enthalpy of formation of $B_4C$/Pd is confirmed. However we cannot tell if Pd and Y form any compound or alloy because even if they do, the binding energy of the alloy peak would be very close to the one of the metal peak. Given the current energy resolution, it is impossible to distinguish alloy and metal from the HAXPES spectra. The decomposition of Pd $2p_{3/2}$ photoemission spectra of other samples brings the same conclusion, thus is not presented here.

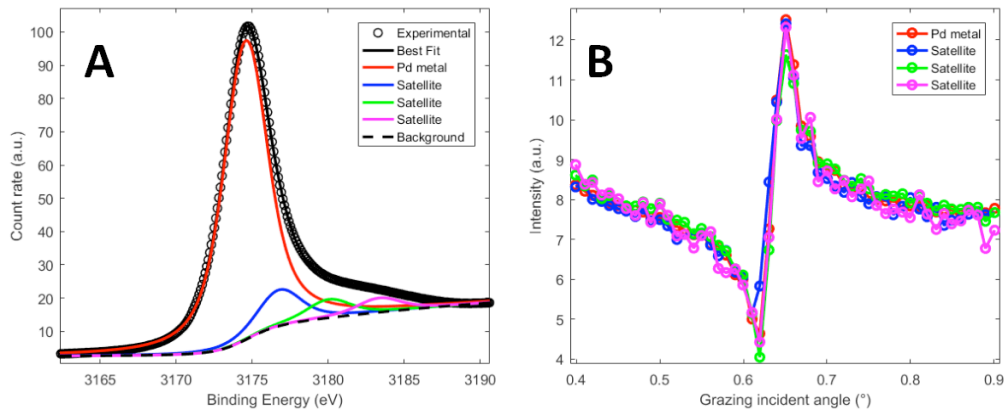

Figure 4. A: decomposition of the sum of all Pd $2p_{3/2}$ photoemission spectra of sample 2 and B: HAXPES-XSW curves of the corresponding components.

Analogous treatment is done for the Y $2p_{3/2}$ core level peak. The experimental spectrum is best fitted using 2 Voigt with 2.00 eV Gaussian width and 1.43 eV Lorentzian width [18]. As presented in Figure 5(A), the major component located at 2078.9 eV in binding energy should be Y metal. The oxidation does not penetrate into the Y layers as previously discussed. The peak which is found at 2081.6 eV stands for the chemical compound of either Y-C or Y-B. To explore the depth distribution of these two chemical states of Y, their HAXPES-XSW curves are depicted for samples 2, 3 and 4 in Figure 5(B)-(D) respectively. First we look at the curves of the sample 3 $[B_4C/Pd/Y]_{40}$ in Figure 5(C). Compared to Y metal, the HAXPES-XSW curve of the compound shifts towards the higher angles, indicating that such compound is located deeper than Y metal in each period. It is then located at the Y-on-$B_4C$ interfaces. On the contrary, for the sample 4 $[Pd/B_4C/Y]_{40}$ in Figure 5(D), we have the angular shift of compound curve towards the lower angles. The compound is then located at a shallower depth than the Y metal, which should be the $B_4C$-on-Y interfaces. The unique appearance of such compound at Y-$B_4C$ interfaces confirms our assumption of its nature: Y-B or Y-C. In the case of sample 2 $[B_4C/Pd/B_4C/Y]_{20}$ we have $B_4C$ barrier layers on both sides of Y layer. In Figure 5(B), we observe an angular shift of the compound curve towards the lower angles. It means that the formation of this kind of Y compound has a preference, or is more active, to happen at $B_4C$-on-Y interfaces than at Y-on-$B_4C$ interfaces. Like the Pd $2p_{3/2}$ spectra, Y $2p_{3/2}$ spectra cannot bring us information on whether there is any chemical compound formed at Pd-Y interfaces because it is not possible to distinguish the alloy peak from the metal peak on the HAXPES spectra. The HAXPES-XSW curves of C 1s and B 1s (not presented) are very noisy due to the low intensity of their photoemission spectra. As a consequence, we cannot tell if the Y compound is Y-B or Y-C. The results obtained by HAXPES-XSW are in line with the values of enthalpy of formation $\Delta H_f$ found in the literature [12].

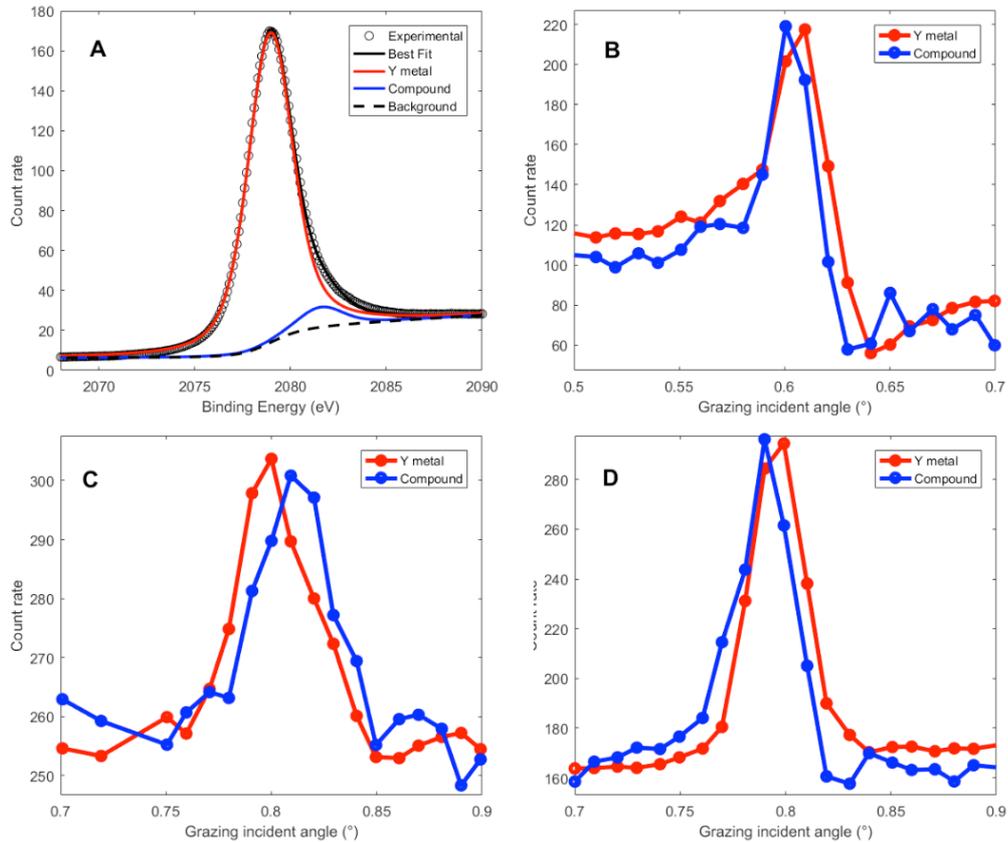

Figure 5. A: decomposition of the sum of all Y $2p_{3/2}$ photoemission spectra of sample 2. HAXPES-XSW curves of the corresponding component peaks of different samples, B: sample 2, C: sample 3, D: sample 4.

The structural parameters in Table 1 determined by GIXR fitting are introduced into the software YXRO [14] in order to calculate the variation of the photoemission intensity as a function of the grazing incident angle. The calculation takes into account of the complex refractive index of each material, the atomic cross section as well the structure of the multilayer. The IMFP of the photoelectrons is calculated by YXRO. Such prediction is presented in Figure 6 for sample 2 where the HAXPES-XSW curves of Pd $2p_{3/2}$, Y $2p_{3/2}$ and B 1s core levels are calculated with an angular range around the first order Bragg angle. As seen in Figure 1(B), when the grazing incident angle is at the Bragg angle (0.63° for sample 2), the anti-nodal plans of the periodic X-ray standing wave field are located at Y-on-$B_4C$ interfaces. Since both the multilayer and the electric field are periodic with an identical period value, the same field distribution can be expected for each period of the stack with the progressive loss of amplitude due to the attenuation of the radiation. The oscillating structure on the

HAXPES-XSW curves is centered close to the Bragg angle, because away from Bragg angle the XSW endures intensity loss due to the reflectance loss. When the incident angle varies through the Bragg angle, the location of the anti-nodal plans moves accordingly toward deeper location from Y layers into $B_4C$ layers then Pd layers. This angular order of the ionization enhancement of the elements is reflected in Figure 6. As the grazing angle varies from low to high values, a rise of the Y $2p_{3/2}$ HAXPES-XSW curve first appears at 0.61°. It is then followed by the rise of the B 1s and Pd $2p_{3/2}$ curves at 0.63° and 0.66° respectively. As the $B_4C$ layers are sandwiched between the Pd and Y layers, the B 1s HAXPES-XSW curve is rather more symmetrical instead of appearing as the "Z" form like the Pd and Y curves.

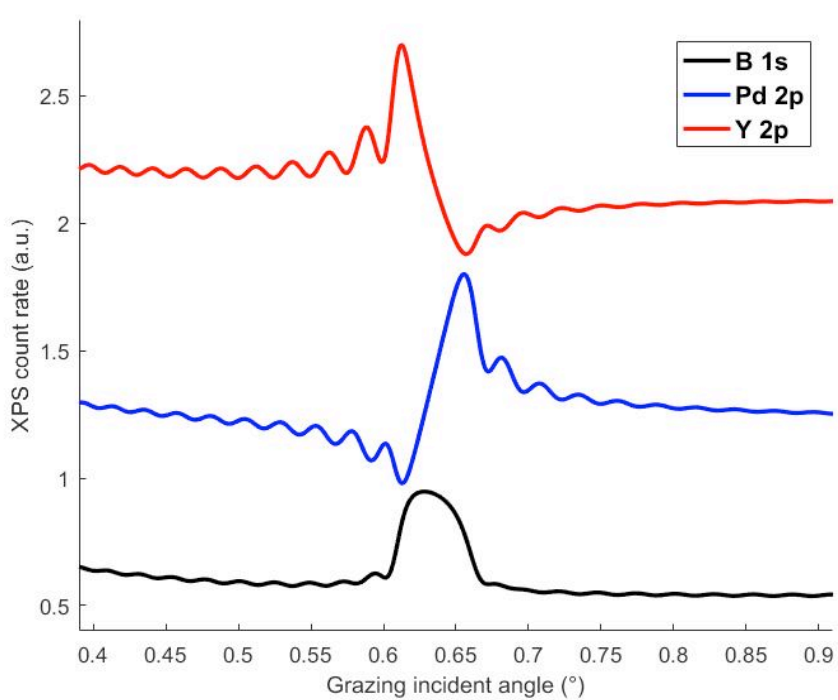

Figure 6. Simulation of the variation of the intensities of core level peaks for sample 2: B 1s in black, Pd $2p_{3/2}$ in blue and Y $2p_{3/2}$ in red. The curves are vertically shifted for the sake of readability.

Secondary oscillation of the calculated curves appears due to the interference between various reflections (so-called Kiessig fringes) resulted from the limited number of periods, which is 20 for sample 2. These simulations are far from what is experimentally observed (Figure 5(B)). This discrepancy is overcome by taking into account of the instrumental angular resolution (0.008°) and the horizontal divergence of the incident photon beam (0.026°). Such broadening can be mathematically simulated by applying a convolution by a gate function onto the simulation curve which is originally calculated with a step of 0.001°. The size of the gate function is then adjusted according to the angular resolution and beam divergence, and the effect is presented in Figure 7 in the case of the Y $2p_{3/2}$

HAXPES-XSW curve. The Kiessig fringes disappear and the adjusted simulation shows a better agreement with the experimental data.

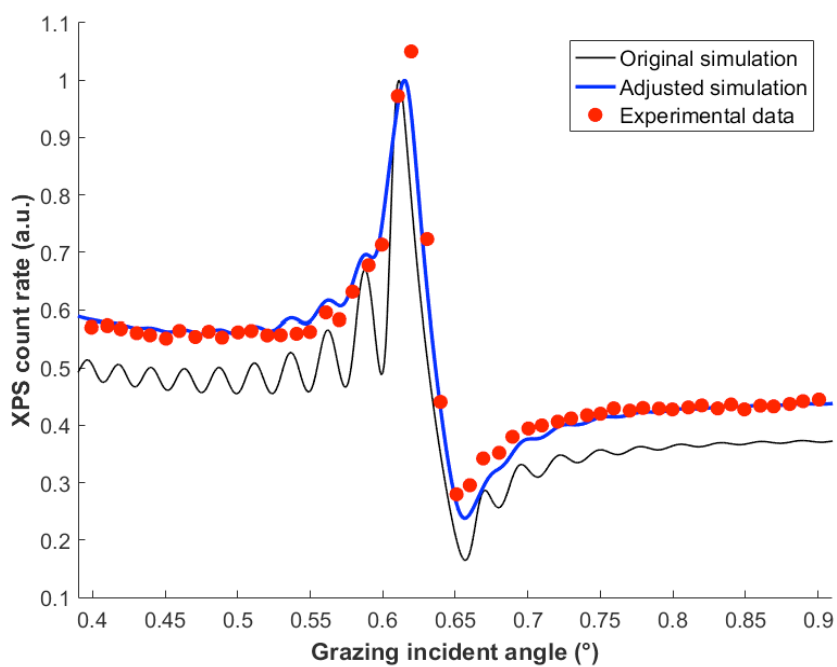

Figure 7. Comparison of the experimental (red dots) and calculated Y $2p_{3/2}$ HAXPES-XSW curves for the sample 2 with (blue line) and without (black line) angular broadening.

Figure 8 presents the HAXPES-XSW curves related to Pd 2p, Y 2p and B 1s peaks of sample 2 and their simulations with the broadening effect considered. The comparison shows a fine agreement indicating that the structure determined by GIXR is reliable as the shapes of these curves tightly correspond to the distribution of each element. For the other samples, the fitting of experimental curves based on the GIXR structural parameters the experimental curves is much less successful. The model used for GIXR fitting may be far too simple to describe the structure of the multilayer, especially concerning the depth distribution of all elements, even all chemical states. The reason that it works fine for sample 2 could be that the $B_4C$ layers on each interface stabilize the multilayer by preventing the interdiffusion of Pd and Y atoms, making the situation relatively simpler compared to other samples. The model used for the GIXR fitting of such 4-layer sample is thus suitable in this case. Unfortunately for the moment we do not possess a fitting process to determine the structure of the multilayer independently.

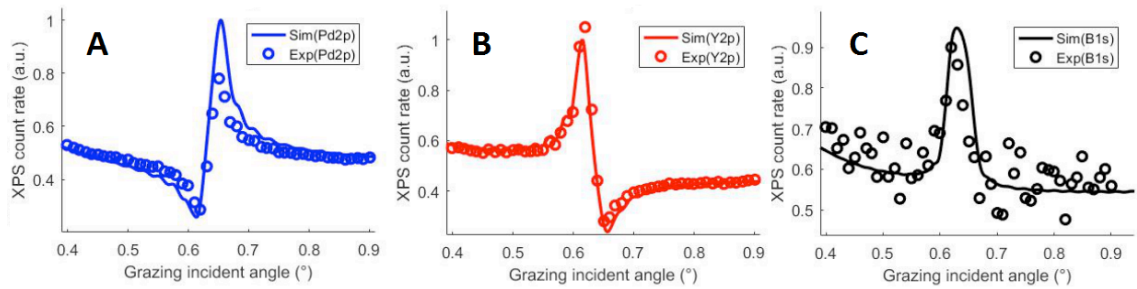

Figure 8. Comparison of experimental and simulated HAXPES-XSW curves for sample 2. A: Pd $2p_{3/2}$, B: Y $2p_{3/2}$ and C: B 1s.

## 4. CONCLUSION

The interdiffusion of the two metals in the Pd/Y system, predicted by calculating the mixing enthalpy using the Miedema model, is clearly seen in the GIXR spectrum of this sample. To study the efficiency of the insertion of $B_4C$ barrier layer at one or both interfaces, $B_4C$/Pd/Y, Pd/$B_4C$/Y and $B_4C$/Pd/$B_4C$/Y multilayers are also considered. The thickness and roughness/interdiffusion of the Pd and Y layers in these series are determined by fitting the experimental GIXR spectra. The interfaces are found to be asymmetrical since interdiffusion is stronger at Y-on-Pd interfaces than at Pd-on-Y interfaces.

The HAXPES-XSW measurements further help explore the nature of the interdiffusion of Pd and Y. The interdiffusion of Pd into the $B_4C$ layers does not form any chemical compound. On the contrary, Y forms chemical compound with either B or C at the Y-$B_4C$ interfaces. The formation of Y compound has a preference, or is more active, to happen at $B_4C$-on-Y interfaces than at Y-on-$B_4C$ interfaces. Such chemical compound can be the reason why the interfaces are stabilized.

Multiple HAXPES-XSW curves corresponding to different elements (or even different core level of one element) are measured. A detailed description of the depth distribution of each element can be obtained by comparing the HAXPES-XSW curves with the calculations using YXRO. However a fitting process, combining both XRR and HAXPES-XSW curves, is in need in order to independently build up a model of the multilayer to determine its structure.

[1] Montcalm, C., Kearney, P. A., Slaughter, J. M., Sullivan, B. T., Chaker, M., Pépin, H. & Falco, C. M. (1996). *Applied Optics*. **35**, 5134–5147. doi:10.1364/AO.35.005134.

[2] Xu, D., Huang, Q., Wang, Y., Li, P., Wen, M., Jonnard, P., Giglia, A., Kozhevnikov, I. V., Wang, K., Zhang, Z. & Wang, Z. (2015). *Optics Express*. **23**, 33018. doi:10.1364/OE.23.033018.

[3] Wu, M.-Y., Ilakovac, V., André, J.-M., Le Guen, K., Giglia, A., Rueff, J.-P., Huang, Q.-S., Wang, Z.-S. & Jonnard, P. (2017). *SPIE Proceedings*. 102350F. doi:10.1117/12.2265630.

[4] Windt, D. L. & Gullikson, E. M. (2015). *Applied Optics*. **54**, 5850. doi:10.1364/AO.54.005850.

[5] Fadley, C. S. (2016). *Hard X-Ray Photoelectron Spectroscopy (HAXPES)*, Edited by J. Woicik , pp. 1–34. Springer International Publishing, Cham. doi:10.1007/978-3-319-24043-5_1.

[6] Fadley, C. S. (2013). *Journal of Electron Spectroscopy and Related Phenomena*. **190**, 165-179. doi:10.1016/j.elspec.2013.06.008.

[7] Batterman, B. W., Gaal, P., Reimann, K., Woerner, M. & Elsaesser, T. (2005). *Optics Letters*. **30**, 2805–2807. doi:10.1364/OL.30.002805.

[8] Wu, M., Burcklen, C., André, J.-M., Guen, K. L., Giglia, A., Koshmak, K., Nannarone, S., Bridou, F., Meltchakov, E., Rossi, S. de, Delmotte, F. & Jonnard, P. (2017). *Optical Engineering*. **56**, 117101. doi:10.1117/1.OE.56.11.117101.

[9] Tu, Y.-C., Yuan, Y.-Y., Le Guen, K., André, J.-M., Zhu, J.-T., Wang, Z.-S., Bridou, F., Giglia, A. & Jonnard, P. (2015). *Journal of Synchrotron Radiation*. **22**, 1419–1425. doi:1419–1425.10.1107/S1600577515016239.

[10] Giglia, A., Mukherjee, S., Mahne, N., Nannarone, S., Jonnard, P., Le Guen, K., Yuan, Y.-Y., André, J.-M., Wang, Z.-S., Li, H.-C. & Zhu, J.-T. (2013). *SPIE Proceedings*. 87770I. doi:10.1117/12.2017252.

[11] Kardellass, S., Selhaoui, N., Iddaoudi, A., Ait Amar, M., Karioui, R. & Bouirden, L. (2013). *MATEC Web of Conferences*. **5**, 04032. doi:10.1051/matecconf/20130504032.

[12] Meschel, S. V. & Kleppa, O. J. (2001). *Journal of Alloys and Compounds*. **321**, 183–200. doi:10.1016/S0925-8388(01)00966-5.

[13] Windt, D. L. (1998). *Computers in Physics*. **12**, 360. doi:10.1063/1.168689.

[14] Yang, S.-H., Gray, A. X., Kaiser, A. M., Mun, B. S., Sell, B. C., Kortright, J. B. & Fadley, C. S. (2013). *Journal of Applied Physics*. **113**, 073513. doi:10.1063/1.4790171.

[15] Céolin, D., Ablett, J. M., Prieur, D., Moreno, T., Rueff, J.-P., Marchenko, T., Journel, L., Guillemin, R., Pilette, B., Marin, T. & Simon, M. (2013). *Journal of Electron Spectroscopy and Related Phenomena*. **190**, 188–192. doi:10.1016/j.elspec.2013.01.006.

[16] Scofield, J. H. (1973). Theoretical photoionization cross sections from 1 to 1500 keV. doi:10.2172/4545040.

[17] Shirley, D. A. (1972). *Physical Review B*. **5**, 4709. doi:10.1103/PhysRevB.5.4709.

[18] Campbell, J. L. & Papp, T. (2001). *Atomic Data and Nuclear Data Tables*. **77**, 1–56. doi:10.1006/adnd.2000.0848.

[19] De Siervo, A., Landers, R., De Castro, S. G. C. & Kleiman, G. G. (1998). *Journal of Electron Spectroscopy and Related Phenomena*. **88**, 429–433. doi:429–433.10.1016/S0368-2048(97)00194-1.